\documentclass[prl,aps,twocolumn,nofootinbib,preprintnumbers]{revtex4}
\usepackage{graphicx,amsmath}

\begin{document}

\def\beq{\begin{eqnarray}}
\def\eeq{\end{eqnarray}}
\newcommand{\gsim}{ \mathop{}_{\textstyle \sim}^{\textstyle >} }
\newcommand{\lsim}{ \mathop{}_{\textstyle \sim}^{\textstyle <} }
\newcommand{\vev}[1]{ \left\langle {#1} \right\rangle }
\newcommand{\bra}[1]{ \langle {#1} | }
\newcommand{\ket}[1]{ | {#1} \rangle }
\newcommand{\EV}{ {\rm eV} }
\newcommand{\KEV}{ {\rm keV} }
\newcommand{\MEV}{ {\rm MeV} }
\newcommand{\GEV}{ {\rm GeV} }
\newcommand{\TEV}{ {\rm TeV} }
\newcommand{\bea}{\begin{eqnarray}}   
\newcommand{\eea}{\end{eqnarray}}
\newcommand{\bear}{\begin{array}}  
\newcommand {\eear}{\end{array}}
\newcommand{\bef}{\begin{figure}}  
\newcommand {\eef}{\end{figure}}
\newcommand{\bec}{\begin{center}}  
\newcommand {\eec}{\end{center}}
\newcommand{\non}{\nonumber}  
\newcommand {\eqn}[1]{\beq {#1}\eeq}
\newcommand{\la}{\left\langle}  
\newcommand{\ra}{\right\rangle}
\newcommand{\ds}{\displaystyle}
\def\SEC#1{Sec.~\ref{#1}}
\def\FIG#1{Fig.~\ref{#1}}
\def\EQ#1{Eq.~(\ref{#1})}
\def\EQS#1{Eqs.~(\ref{#1})}
\def\GEV#1{10^{#1}{\rm\,GeV}}
\def\MEV#1{10^{#1}{\rm\,MeV}}
\def\KEV#1{10^{#1}{\rm\,keV}}
\def\lrf#1#2{ \left(\frac{#1}{#2}\right)}
\def\lrfp#1#2#3{ \left(\frac{#1}{#2} \right)^{#3}}

%


\preprint{UT-14-03}
\title{Suppressed Non-Gaussianity in the Curvaton Model}
\renewcommand{\thefootnote}{\alph{footnote}}

\author{
Kyohei Mukaida$^{a}$,
Kazunori Nakayama$^{a,b}$,
and 
Masahiro Takimoto$^{a}$}

\affiliation{
 $^a$Department of Physics, University of Tokyo, Tokyo 113-0033, Japan\\
 $^b$Kavli Institute for the Physics and Mathematics of the Universe, Todai Institute for Advanced Study, University of Tokyo, Kashiwa 277-8583, Japan
  }

\begin{abstract}

We show that the local type non-Gaussianity in a class of curvaton models is suppressed,
i.e. the non-linearity parameters $f_{\rm NL}$ and those related with higher order statistics can be at most $O(1)$,
even if the curvaton energy density is subdominant at the decay.
This situation is naturally realized in a very simple curvaton potential with quadratic term plus quartic term.

\end{abstract}

\maketitle

In the era of precision cosmology after the Planck satellite~\cite{Ade:2013zuv},
the idea of inflation in the very early universe~\cite{Guth:1980zm} seems to be firmly confirmed.
Inflation naturally gives the origin of primordial curvature perturbation, 
which is the seed of all the observed rich structure of the present universe.

In the inflationary universe scenario, there are two main possibilities as a source of the curvature perturbation.
One is the inflaton: the quantum fluctuation of the scalar field which causes inflationary expansion
can be the origin of the primordial curvature perturbation.
In the curvaton scenario, on the other hand, the curvature perturbation is generated by the quantum fluctuation of a light scalar field
other than the inflaton, called curvaton~\cite{Mollerach:1989hu,Linde:1996gt,Enqvist:2001zp,Lyth:2001nq,Moroi:2001ct}.
The curvaton is assumed to have a negligible energy fraction during inflation,
but later it tends to (nearly) dominate the energy density of the universe before it decays into radiation.
Currently the curvaton scenario is still viable in light of the Planck result.

One of the characteristic features of the curvaton scenario is that it can produce large non-Gaussianity,
often parameterized by $f_{\rm NL}$~\cite{Komatsu:2001rj}.
If the curvaton energy density is subdominant at the decay, $f_{\rm NL}$ is enhanced as $f_{\rm NL} \sim 1/R$
where $R$ is roughly the energy fraction of the curvaton at the decay~\cite{Lyth:2001nq,Lyth:2002my}.
The Planck constraint reads $f_{\rm NL}=2.7\pm5.8~~ (68\%~ \text{C.L.})$~\cite{Ade:2013ydc}.
Thus if the curvaton scenario is true, the curvaton energy fraction $R$ cannot be so small
in order to avoid too large non-Gaussianity.

In this letter we show that this naive expectation breaks down in a very simple and natural curvaton model.
Actually the local type non-Gaussianity can naturally be suppressed so that $f_{\rm NL}$ can be at most $\mathcal O(1)$
independently of the curvaton energy fraction.
Moreover, any non-linearity parameter related to higher order statistics, such as $g_{\rm NL}$, can also be at most $\mathcal O(1)$.
A sufficient condition is that the curvaton begins to oscillate in the quartic potential in the inflaton-oscillation dominated era
and decays when the potential is dominated by the quadratic one.\\

Let us consider the curvaton with the canonical kinetic term and the following potential
\begin{equation}
	V(\phi) = \frac{1}{2}m^2\phi^2 + \frac{\lambda}{n} \phi^n,
	\label{pot}
\end{equation}
with $n>2$.\footnote{
	Following calculations are formally applicable also for $n=2$,
	if $\phi_c$ is taken anywhere in the potential.
}
The cosmological scenario is as follows.
The curvaton $\phi$ is placed at $\phi_i$ $(\gg \phi_c)$ during inflation, 
where $\phi_c = (m^2/\lambda)^{1/(n-2)}$, and begins to oscillate at $H=H_{\rm os}$,
where $H$ is the Hubble parameter, during the inflaton oscillation dominated era.
After that, the amplitude of the coherent oscillation, $\tilde\phi$, decreases as $\tilde\phi \propto a(t)^{-6/(n+2)}$
with $a(t)$ being the scale factor of the universe.
Then the curvaton potential becomes dominated by the quadratic one around $H=H_c$.
Inflaton decays at $H\simeq\Gamma_{\rm inf}$, with $\Gamma_{\rm inf}$ being the inflaton decay rate, 
after that the radiation dominated era begins.
At this stage, we assume that the curvaton energy density is still negligible compared with that of radiation.
Finally, the curvaton decays at $H\simeq\Gamma_\phi$ with $\Gamma_\phi$ being the curvaton decay width
and the dominant curvature perturbation is generated.
The following results do not depend on whether $H_c > \Gamma_{\rm inf}$ or $H_c < \Gamma_{\rm inf}$,
but hereafter we assume $H_c > \Gamma_{\rm inf}$ for concreteness.

Here it should be noticed that there are three sources of curvature perturbation.
One is the fluctuation of $\rho_\phi$ itself, another is the modulation of the oscillation epoch $H_{\rm os}$ due to the $\phi^n$ term in the potential~\cite{Enqvist:2005pg,Kawasaki:2008mc,Kawasaki:2011pd}
and the other is the modulation of $H_c$~\cite{Byrnes:2011gh,Mukaida:2014yia}.
This last contribution is crucial for the discussion below and this effect has been neglected in the most previous literatures.

We make use of the $\delta N$ formalism to calculate the non-linear curvature perturbation $\zeta$~\cite{Starobinsky:1986fxa,Sasaki:1995aw,Sasaki:1998ug,Lyth:2004gb}
\begin{equation}
	\zeta = N(\vec x) - \bar N,
\end{equation}
where $N(\vec x)$ is the e-folding number between the final uniform density slice and the initial spatially flat slice,
and $\bar N$ is that of the background value.
We take the final uniform density slice to be the curvaton decay surface: $3H(\vec x)=\Gamma_\phi = {\rm const}$.
On our purpose, the initial slice can be taken at arbitrary time as long as the curvaton energy density is negligible
and the curvaton is the only source of the curvature perturbation.
Hence we take it to be the reheating surface $3H=\Gamma_{\rm inf}$ just for simplicity.

The curvaton energy density at $3H=\Gamma_{\rm inf}$ is evaluated as
\begin{equation}
	\rho_\phi^{\rm (reh)}(\vec x) = \rho_{\phi_i}(\vec x) \left( \frac{H_c(\vec x)}{H_{\rm os}(\vec x)} \right)^{\frac{4n}{n+2}} 
	\left( \frac{\Gamma_{\rm inf}/3}{H_{c}(\vec x)} \right)^{2}. 
\end{equation}
Note that $ \rho_{\phi_i}(\vec x) \propto \phi_i^n$, $H_{\rm os} \propto \phi_i^{(n-2)/2}$ and $H_c \propto \phi_i^{(n-6)/4}$.
Thus it is schematically rewritten as
\begin{equation}
	\rho_\phi^{\rm (reh)}(\vec x) \equiv B \phi_i^p(\vec x),~~~~~p=\frac{6-n}{2}. \label{rho_reh}
\end{equation}
Therefore, the curvaton energy density at its decay $H=\Gamma_\phi$ is given by
\begin{equation}
	\rho_\phi^{\rm (dec)}(\vec x) =  \rho_\phi^{\rm (reh)}(\vec x)\left( \frac{a(t_{\rm reh})}{a(t_{\rm dec})} \right)^3
	= B \phi_i^p(\vec x) y^3,
	\label{rho_dec}
\end{equation}
where $a(t_{\rm reh})$ and $a(t_{\rm dec})$ stand for the scale factor at the inflaton decay and the curvaton decay, respectively,
and $y\equiv a(t_{\rm reh})/a(t_{\rm dec})$.

On the other hand, radiation energy density at the curvaton decay is evaluated as
\begin{equation}
	\rho_r^{\rm (dec)}(\vec x) =  \rho_r^{\rm (reh)} \left( \frac{a(t_{\rm reh})}{a(t_{\rm dec})} \right)^4 \equiv A y^4.
\end{equation}
Here it should be noticed that $\rho_r^{\rm (reh)}$ does not depend on $\vec x$ because of the assumption that the
curvaton energy density is negligible at the inflaton decay.

Then we have the following condition at the curvaton decay surface~\cite{Sasaki:2006kq}:
\begin{equation}
	A y^4 + B \phi_i^p y^3 = \frac{1}{3}\Gamma_\phi^2 M_P^2 = {\rm const}.  \label{curvdec}
\end{equation}
By differentiating this equation with respect to $\phi_i$, we obtain
\begin{equation}
	4A y^{(1)} + pB\phi_i^{p-1} + 3B \phi_i^p \frac{y^{(1)}}{y} = 0,  \label{y1}
\end{equation}
where the superscript ${(n)}$ means the $n$-th derivative with respect to $\phi_i$.
By noting $N(\vec x) = -\ln y$, we obtain
\begin{equation}
	N^{(1)} = -\frac{y^{(1)}}{y} = \frac{pR}{3\phi_i},
\end{equation}
where 
\begin{equation}
	R \equiv \frac{3\rho_\phi^{\rm (dec)} }{3\rho_\phi^{\rm (dec)} + 4\rho_r^{\rm (dec)}}.
\end{equation}
The curvature perturbation is given by $\zeta \simeq N^{(1)} \delta\phi_i \simeq pR H_{\rm inf}/(6\pi \phi_i)$,
where $H_{\rm inf}$ is the Hubble scale during inflation.
Observationally we have $\zeta \simeq 5\times 10^{-5}$~\cite{Ade:2013zuv}
and it fixes the inflation scale $H_{\rm inf}$ for given parameter sets.
Similarly, by differentiating Eq.~(\ref{y1}) with $\phi_i$ several times, we can obtain
\begin{equation}
	N^{(2)} = \frac{p R}{9\phi_i^2}\left[ 3(p-1) - 2p R - p R^2 \right].  \label{N2}
\end{equation}
and
\begin{equation}
\begin{split}
	N^{(3)} =& \frac{p R}{27\phi_i^3} \left[ 9(p-1)(p-2) + 18p(1-p)R \right.\\
	&\left. + p(9-4p)R^2 + 10p^2 R^3 + 3p^2 R^4 \right].  \label{N3}
\end{split}
\end{equation}
They reduce to the ordinary formulae for the curvature perturbation from the curvaton for $n=2$ (see e.g. Refs.~\cite{Sasaki:2006kq,Lyth:2005fi}).

\begin{figure}[t!!]
\begin{center}
\includegraphics[scale=0.5]{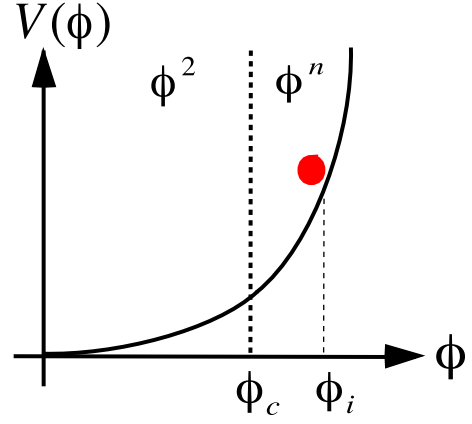}
\caption{ 
	Schematic picture of the curvaton potential (\ref{pot}).
	The initial position during inflation is $\phi_i$, which is larger than $\phi_c$ where the dominant term in the potential changes.
}
\label{fig:pot}
\end{center}
\end{figure}

An interesting situation appears in the case of $n=4$ or $p=1$.\footnote{
	Another interesting case is $n=6$ or $p=0$. 
	In this case the curvature perturbation vanishes  at the non-linear level as is clear from Eq.~(\ref{curvdec}).
	This is because the modulation of the oscillation epoch, encoded in $H_{\rm os}$,
	cancels the fluctuation of $\rho_{\phi_i}$~\cite{Dimopoulos:2003ss}.
}
Note that the quartic term arises as a Coleman-Weinberg correction~\cite{Coleman:1973jx} if $\phi$ couples to
another scalar field $\chi$ as $V \propto \phi^2\chi^2$.
In this case, it is seen that the first term in Eq.~(\ref{N2}) vanishes.
The non-linearity parameter $f_{\rm NL}$ is given by~\cite{Lyth:2005fi}
\begin{equation}
	\frac{6}{5}f_{\rm NL} = \frac{N^{(2)}}{(N^{(1)})^2} = -2-R.
\end{equation}
There is no enhancement like $f_{\rm NL} \sim 1/R$ as opposed to the ordinary consideration.
Moreover, the non-linearity parameter which characterizes the trispectrum is obtained as
\begin{equation}
	\frac{54}{25}g_{\rm NL} = \frac{N^{(3)}}{(N^{(1)})^3} = 5 + 10R + 3R^2.
\end{equation}
Both the $1/R$ and $1/R^2$ enhancement terms vanish.
Actually we can formally show that all the non-linearity parameters of the form
$\sim N^{(n)}/(N^{(1)})^n$ do not contain terms with negative powers of $R$. See Appendix for details.

What is crucial in this argument is that $\rho_\phi^{\rm (dec)}$ [Eq.~(\ref{rho_dec})] linearly scales with $\phi_i$ for $p=1$,
except for the small $\phi_i$ dependence in $y$.
This should be contrasted to the ordinary curvaton model in which $\rho_\phi^{\rm (dec)} \propto \phi_i^2$, corresponding to $p=2$
(or $n=2$).
In such a case, the curvature perturbation is roughly given by the fluctuation of the curvaton energy density,
$\zeta \sim R(\delta \rho_\phi^{\rm (dec)}/\rho_\phi^{\rm (dec)}) = R(2\delta_{\phi_i} + \delta_{\phi_i}^2)$
with $\delta_{\phi_i} \equiv \delta\phi_i/\phi_i$.
This second term was a source of large non-Gaussianity: $f_{\rm NL} \sim 1/R$ for $R\ll 1$.
In our case, for $p=1$, this second term does not exist since  $\rho_\phi^{\rm (dec)}\propto \phi_i$.
All the non-linearity arises from the small $\phi_i$ dependence of $y$, 
which however does not generate non-Gaussianity enhanced with negative powers of $R$.\\

\begin{figure}[t!!]
\begin{center}
\includegraphics[scale=0.9]{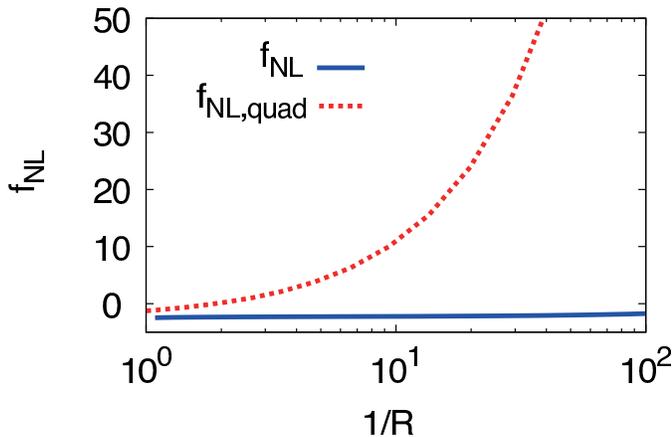}
\caption{ 
	Numerical result of $f_{\rm NL}$ as a function of $1/R$ for the potential (\ref{pot}) with $n=4$ (blue solid).
	For comparison, $f_{\rm NL}$ with simple quadratic potential ($f_{\rm NL, quad}$, red dashed) is also plotted.
}
\label{fig:fNL}
\end{center}
\end{figure}

We have performed numerical calculations to confirm the arguments so far.
We have numerically solved following set of equations:
\begin{gather}
	\dot\rho_{\rm inf} + 3H\rho_{\rm inf} = -\Gamma_{\rm inf}\rho_{\rm inf},\\
	\dot\rho_{r} + 4H\rho_{r} = \Gamma_{\rm inf}\rho_{\rm inf} + \Gamma_\phi\rho_\phi, \\
	\ddot\phi + (3H+\Gamma_\phi)\dot\phi + \frac{\partial V}{\partial \phi} = 0,\\
	H^2 = \frac{1}{3M_P^2}(\rho_{\rm inf} + \rho_{r} + \rho_{\phi}),\\
	\rho_\phi = \frac{1}{2}\dot\phi^2 + V,
\end{gather}
where $\rho_{\rm inf}$ denotes the inflaton energy density and $M_P$ is the reduced Planck scale.
The result for $n=4$ is shown as a blue solid line in Fig.~\ref{fig:fNL}.
Parameters are chosen as $(m,\lambda,T_{\rm R},\phi_i)=(10^3\,{\rm GeV},10^{-18},
10^9\,{\rm GeV},10^{15}\,{\rm GeV})$ with $T_{\rm R}$ being the reheating temperature of the universe:
\begin{align}
	\Gamma_{\rm inf}^2=\frac{g_{\ast}\pi^2T_{\rm R}^4}{10M^2_{P}},
\end{align}
where $g_{\ast}=106.75$ is the effective relativistic degrees of freedom.
Actually, with this set up, the curvaton starts to oscillate and the effective potential is shifted to the quadratic one in the inflaton dominated era. By varying $\Gamma_\phi$, we scan wide range of $R$.
Then we evaluate the e-folding number from the initial time to the uniform Hubble slice, taken well after the curvaton decay,
for various initial condition $\phi_i$, to obtain $N^{(1)}$ and $N^{(2)}$.
It is seen that even for small $R$, $f_{\rm NL}$ remains almost constant of $\mathcal O(1)$ as expected.
For comparison, we have also shown numerical result of $f_{\rm NL}$ for the simple quadratic potential $V = m^2\phi^2/2$
by the red dashed line, which shows the enhancement like $f_{\rm NL} \sim 1/R$ for small $R$.
\\

So far we have assumed that the curvaton begins to oscillate in the inflaton-oscillation dominated era.
If the curvaton begins to oscillate in the radiation dominated era after the reheating,
we obtain similar results except that $p$ is replaced by $p=(10-n)/4$.
Hence the non-Gaussianity is suppressed for $n=6$, as studied in Ref.~\cite{Byrnes:2011gh}, 
and the curvature perturbation vanishes for $n=10$.
\\

To summarize, we have shown that the local type non-Gaussianity can naturally be suppressed in a simple curvaton model like (\ref{pot}) with $n=4$.
If the curvaton potential is initially dominated by $\phi^4$ term and begins to oscillate in the inflaton-dominated era,
all the non-linearity parameters do not contain terms proportional to negative powers of $R$.
It says that the curvaton non-Gaussianity cannot be large even in the simplest setup, contrary to the common wisdom.
If the curvaton potential contains more than two polynomials, results would be more complicated.
In any case, calculations of non-linearity parameters in various curvaton potentials and cosmological evolution scenarios
may need to be reexamined, in particular taking the effect of modulation of $H_c$ into account.
We emphasize that in some cases, as studied in this letter, it happens that $f_{\rm NL}$ cannot have enhancement like $\sim 1/R$
or even the curvature perturbation itself vanishes.

\section*{Note added} \label{sec:note}

Very recently, after the submission of this manuscript, the BICEP2 experiment reported the detection of B-mode polarization
with the tensor-to-scalar ratio $r=0.20^{+0.07}_{-0.05}$~\cite{Ade:2014xna}.
It indicates the high-scale inflation of $H_{\rm inf} \simeq 10^{14}\,$GeV
and it is consistent with the chaotic inflation scenario with quadratic potential~\cite{Linde:1983gd}.
If this is true, the inflaton can naturally produce right amount of the density perturbation.
Still, however, the curvaton may contribute to a part of the density perturbation~\cite{Byrnes:2014xua,Sloth:2014sga}
with potentially too large non-Gaussianity~\cite{Fujita:2014iaa}.
Interestingly, it is pointed out that the sizable anti-correlated isocurvature perturbation, which may be produced in the curvaton model,
relaxes the tension between BICEP2 and Planck results~\cite{Kawasaki:2014lqa}.
Our mechanism to suppress the non-Gaussianity is useful in such a mixed inflaton-curvaton scenario.
The non-linearity parameter in the mixed inflaton-curvaton scenario is given by
\begin{equation}
	\frac{6}{5}f_{\rm NL} =\left(\frac{\mathcal E}{1+\mathcal E}\right)^2 \frac{3(p-1)-2pR-pR^2}{pR},
\end{equation}
where $\mathcal E$ denotes the ratio of the curvature perturbation from the curvaton to that from the inflaton:
\begin{equation}
	\mathcal E \equiv \left( \frac{N^{(1)}}{N_I} \right)^2.
\end{equation}
Here we have expanded the curvature perturbation as $\zeta = N_I\delta I + N^{(1)} \delta \phi + \dots$
with $I$ denoting the inflaton field.
Numerically, we have
\begin{equation}
	\mathcal E \sim 10^{-2} p^2R^2 \left( \frac{10^{18}\,{\rm GeV}}{\phi_i} \right)^2. 
\end{equation}
Thus, for $p=2$ or $n=2$, the non-linearity parameter can be too large if $R\ll 1$ for fixed $\mathcal E \sim \mathcal O(0.1)$
as explicitly shown in Ref.~\cite{Fujita:2014iaa} recently.
On the other hand, for $p=1$ or $n=4$, there is no $1/R$ enhancement (for fixed $\mathcal E$) 
and hence such a constraint is safely avoided.
The same is true for other non-linearity parameters such as $\tau_{\rm NL}, g_{\rm NL}$ and so on.

\section*{Acknowledgments}

This work is supported by Grant-in-Aid for Scientific
research from the Ministry of Education, Science, Sports, and Culture
(MEXT), Japan, No.\ 21111006 (K.N.), and No.\ 22244030 (K.N.).
The work of K.M. and M.T. are supported in part by JSPS Research Fellowships
for Young Scientists.
The work of M.T. is supported by the Program for Leading Graduate
Schools, MEXT, Japan.

\appendix
\section{Formal proof of suppressed non-linearity parameters} \label{sec:app}

In this Appendix we give a formal proof of the vanishing terms proportional to negative powers of $R$ in the expression of
non-linearity parameters for $n=4$ $(p=1)$.
We define
\begin{equation}
	F[\ell] \equiv \frac{N^{(\ell)}}{(N^{(1)})^\ell} = \frac{\partial^\ell N (\phi_i) / \partial \phi_i^\ell}{[\partial N(\phi_i) / \partial \phi_i ]^\ell},
\end{equation}
which is related to non-linearity parameters characterizing $\ell$-point correlation function of the curvature perturbation.

By differentiating Eq.~(\ref{curvdec}) with respect to $\phi_i$ repeatedly, we find
\begin{equation}
	4A y^{(\ell)} - 3B\left[ (\ell-1)N^{(\ell-1)}+ \phi_i N^{(\ell)}  \right] = 0,
\end{equation}
for $\ell \geq 2$.
On the other hand, $y^{(\ell)}$ and $N^{(\ell)}$ are related through
\begin{equation}
	\frac{y^{(\ell)}}{y} = - N^{(\ell)} + G[\ell],
\end{equation}
where $G[\ell]$ consists of the products of the form $N^{(\ell_i)} \cdots N^{(\ell_k)}$ with $\ell_i +\cdots + \ell_k = \ell$ and $\ell_i \leq \ell-1$.
Therefore, we obtain
\begin{equation}
	F[\ell] = (1-R) \frac{G[\ell]}{ (N^{(1)})^\ell} -3(\ell-1)F[\ell-1].
\end{equation}
The first term can be written by the products of the form $F[\ell_i] \cdots F[\ell_k]$ with $\ell_i +\cdots + \ell_k = \ell$ and $\ell_i \leq \ell-1$.
Therefore, $F[\ell]$ is always expressed in terms of $F[\ell_i]$ with $\ell_i \leq \ell-1$ times numerical constants containing only positive power of $R$.
Since $F[1] = 1$, all $F[\ell]$ with $\ell >1$ also do not contain negative powers of $R$.



\end{document}